  \magnification=\magstep1
 \settabs 18 \columns
\hsize=16truecm

 \input epsf

\def\b{\bigskip}
\def\bb{\bigskip\bigskip}

\def\no{\noindent}
\def\r{\rightline}
\def\ce{\centerline}
\def\ve{\vfill\eject}

\def\r{\rightline}

\def\harr#1#2{\smash{\mathop{\hbox to .25 in{\rightarrowfill}}
  \limits^{\scriptstyle#1}_{\scriptstyle#2}}}

\def\R{{\cal R}}

\def\today{\ifcase\month\or January\or February\or March\or April\or
May\or June\or July\or
August\or September\or October\or November\or  December\fi
\space\number\day, \number\year }

\r {\today, Vs3}

\bb\bb\bb




\def\e{\rm e}

\def\p{\partial}

\def\sqr#1#2{{\vcenter{\vbox{\hrule height.#2pt
\hbox{\vrule width.#2pt height#2pt \kern#2pt
\vrule width.#2pt}
\hrule height.#2pt}}}}

  \def\1/2{{\scriptstyle{1\over 2}}}
  \def\a/2{{\scriptstyle{3\over 2}}}
  \def\5/2{{\scriptstyle{5\over 2}}}
  \def\7/2{{\scriptstyle{7\over 2}}}
  \def\3/4{{\scriptstyle{3\over 4}}}

\font\steptwo=cmb10 scaled\magstep2


\def\picture #1 by #2 (#3){
  \vbox to #2{
    \hrule width #1 height 0pt depth 0pt
    \vfill
    \special{picture #3} 
    }
  }

\def\scaledpicture #1 by #2 (#3 scaled #4){{
  \dimen0=#1 \dimen1=#2
  \divide\dimen0 by 1000 \multiply\dimen0 by #4
  \divide\dimen1 by 1000 \multiply\dimen1 by #4
  \picture \dimen0 by \dimen1 (#3 scaled #4)}
  }

%
%

\def\sqr#1#2{{\vcenter{\vbox{\hrule height.#2pt
\hbox{\vrule width.#2pt height#2pt \kern#2pt
\vrule width.#2pt}
\hrule height.#2pt}}}}

\def \r{\rightarrow}

\b

\vskip-2
cm
\ce{\steptwo An equation of state for dark matter in the Milky Way}

\b
\ce{C. Fr\o nsdal and T. J. Wilcox}
\b
\ce{\it Department of Physics and Astronomy, University of California Los Angeles}
\b
\no{\it ABSTRACT}  ~Dark matter,  believed to be present in many galaxies,
is interpreted as a hydrodynamical system in interaction with the gravitational field and with nothing else.   The  gravitational field of our Galaxy can be inferred from observation of orbital velocities of the visible stars,
in a first approximation in which the field is taken to be  due to the distribution of dark matter only.
An equation of state is determined by  the  gravitational field via the equations of motion.

To arrive at an estimate of the distribution of dark matter in our galaxy, and simultaneously learn something about the gravitational field in the inner regions,
the following strategy was adopted:  1. The observed rotation curves suggest an
expression for the newtonian potential, valid in the outer region.  2. The assumption of a quasi stationary, spherically symmetric distribution of dark matter then leads to a unique equation of state. 3. This equation of state is assumed to be valid all the way to the center (though of course the newtonian approximation is not).  4. Using this equation of state, together with Einstein's equations and the relativistic hydrostatic condition, we
calculate the metric and the matter density throughout the galaxy. The solutions are
regular all the way to the center; there is no indication of a structure of the type of
a Black Hole.

The equation of state that is thus determined experimentally is of the type used by Chandrasekhar and others for the 
degenerate Fermi gas. In the approximation of weak fields the associated "sinh-Emden" equation,  $\Delta \mu = a \sinh^4\mu$ has a global, nonsingular solution.

    \b

\no{\bf  I. Introduction}

One of the enduring problems of astrophysics is to place an upper limit 
on the mass of a star. Let us agree from the outset that the  observed,
gravitational  ``mass" $M$
 of a spherically symmetric object is a limiting value or asymptotic value of the function  $M(r)$ that appears in the  asymptotic Schwartzschild metric,
$$
g_{00}(r) = c^2
(1-{2M(r)G\over r}).\eqno(1.1)
$$
A \underbar{ locally observed mass} is defined in a region where this function is slowly varying.
The Great Attractor near the center of the Milky Way is observed, at a distance from the center of around $ 10^{16}$ cm, to have a value
for $M$ that is several million solar masses  (Ghez 2008). This is several orders of magnitude greater than ``reasonable" physical models (Hartle 1978).
The observation has been interpreted as being the effect of a black hole, with the
horizon at about $10^{12}$ cm from the center.

 Analysis of the distribution of velocities of orbiting stars show that   the newtonian potential cannot be attributed to visible sources;  the locally observed mass increases far too rapidly with the distance from the center.   (  Viollier et al,1993,
 Bilic et al 1998, Munyaneza et al, 1998, Gobar et al 2006,  Ahmed et al 2009, 
Genzel et al 2010).
 Both problems can be qualitatively explained in terms of  `dark matter',  the high
 value of $M(r)$ because the equation of state of dark matter is unknown and not subject to the physical constraints of known forms of matter, the unexpected variation of $M(r)$ with
 distance because dark matter may be present in regions that appear to be empty.

All  that is known about dark matter is that it does not interact directly with ordinary matter. It does not interact with electrodynamics,   it is not in thermal equilibrium with ordinary matter or with radiation; the temperature is therefore not observable and in consequence it is not defined.    This puts the theoretician in the same position as the one that he confronts in hydrodynamics when the temperature is eliminated from the theory by means of the
ideal gas equation and the equation of state reduces to  a relation between
density and pressure. The free energy density $f(T,\rho)$  is a function of density alone, the entropy density  $s = \p f/\p T$  is zero and the pressure is
$$
p = \rho{d\over d \rho}f-f.\eqno(1.2)
$$
The system is thus determined by the expression chosen for the function $f(\rho)$;   this expression, or the inferred relation between pressure and density,  is the equation of state.  

A simple, analytic expression for the newtonian
potential   accounts for  the main features of the rotation curves of the outer region of
our Galaxy. In this domain it is not necessary to invoke General Relativity,
accordingly the equations of newtonian gravity are used to determine the equation of state. 
This equation of state turns out to have an interesting relation to equations of
state proposed long ago in connection with the degenerate Fermi gas. 
The ``sinh-Emden equation" for the potential
$$
\Delta \mu = a \sinh^4\mu,~~~a~{\rm constant},
$$
 is solved exactly. 
 
 Once an equation of state has been determined it is natural to assume that it is valid everywhere, but this assumption is contradicted by observation;   it predicts a velocity for the inner orbiters of the Milky Way  that is too slow by almost one order of magnitude.  We propose to interpret this as evidence for 2 types of Dark Matter
 or, alternatively as a change of phase.

 A most fascinating problem is the interpretation of the heavy object in the center of the Galaxy.  Attempts to interpret the data in terms of a Black Hole are interesting
but if taken literally  lack experimental support since there is no evidence as yet of the existence of an event horizon.The Schwarzschild Black Hole is a solution of Einstein's equations FOR EMPTY SPACE. \footnote * {That is, a space that  is empty in the region outside the horizon.}
We shall show that there are well behaved solutions of Einstein's equations 
that account for the existence of very large concentrations of galactic masses and that do not have a horizon.
 
This paper  presents  a simple equation of state that can account for 
large masses and for  strong gravitational fields in the inner region of galaxies.    We have  used  the same type of equation of state that was used for the outer region
of the Milky Way, exploring  a wide range of the values of the parameters.  

 Since a Schwarzschild Black Hole metric is a solution for empty space one   expects  that the actual metric, given
the presence of matter, can be quite different. We have at hand a family
of models (equations of state) for the description of an object with adjustable mass
and  size. As the size is reduced the outer,  nearly empty region in which the Schwarzschild expression for the metric is approximately valid,   extends  further inwards, eventually so far as to reach a distance from the center of the order of the Schwarzschild radius, at which point a horizon might  be expected to appear. 

But no horizon was found, even when the size of the galaxy  was   reduced by
4 orders of magnitude, the mass fixed.   The solutions are regular all the way to the center and no inner structure appears.
 
\b
{\it Notation and some data.} The radius of the Milky Way is about $R = 10^{23} cm$ and the mass is about $2MG= 2\times 10^{17} cm$.  This measure of the mass of a star, in units of centimeters, is the value of the nominal Schwarzschild radius. The innermost, observed satellite has a 
nearest approach of  about $2\times 10^{15}cm$ and it moves in  the newtonian field of a mass of about  $2MG = 10^{12} cm $. The cgs system is used throughout; $1 kpc = 3\times 10^{21} cm$ and  $8\pi G = 1.863\times 10^{-27} cm/g$.
\b

\ce{\bf  Summary}

Section 2.  Knowledge of the equation of state gives information about the nature of Dark Matter that is not available by other means. To properly interpret the evidence we begin with a discussion of equations of state in the context of General Relativity.
The only thing that is known about  the nature of Black Matter is that it is difficult to obtain information about it; we interpret this fact as evidence that it has 
very little to tell us, that the entropy is negligible. With this insight  
the observed orbital velocities can be directly related to an equation of state for Dark Matter.
This equation of state  is   reminiscent of one proposed long ago for a degenerate Fermi gas.  

Sections 3 and 4. Our model for  the gravitational potential in the outer region, in the approximation in which the contribution of visible matter is ignored, is
$$
{2MG\over r} =-\phi(r)  = k \ln{r+b\over  r},~~ k,b~{\rm constants}.
$$
The equations of motion include the hydrostatic condition in integrated form,
$$
{c^2\over 2}\phi  = {d f\over d\rho},
$$
where $f$ is the free energy  density.  Einstein's equations give  a  unique equation of state represented parameterically as follows
$$
f(\rho)= B\psi\sinh^4\psi-p,~~\rho= A \sinh^4{\psi},~~ p =B\int \sinh^4\psi \,d\psi,
$$
 $A$ and $B$  constants.   We solve the relativistic  equations of motion using this equation of state to obtain the gravitational metric and the density distribution of the Galaxy. The parameters $b$ and $k$ (there are no others) are adjusted to get a qualitative fit to the total mass, the extent and the observed rotation curve of the Galaxy. 
 
 Section 5. Using the same equation of state, and  the full structure  of General Relativity, we explore the inner region of a star or a galaxy that consists of Dark Matter.  Using the same equation of state, with the parameters $b$ (the size)
 and $kb$ (the mass)  we solve the equations of General Relativity numerically.
 Fine tuning the initial values (at  a large distance) we run the computer calculations all the way between a distance of 30 kiloparsecsand the center, literally to a distance of 1 cm from the center.  The main conclusion is that the presence of
 the matter prevents the appearance of a horizon.
 
 Section 6. Conclusions and suggestions

 \bb 
 
  \no{\bf 2. On equations of state in stellar models}

Theories of stellar structure and evolution must postulate some properties of
the matter that makes up the star. To be precise, what is needed  is a relation between
density and pressure. In the early models of Lane (1870), Ritter (1870-1880), Emden (1907) and Eddington (1926) the polytropic relation was used,
$$
p =a \rho^\gamma,~~\gamma=1+{1\over n},\eqno(2.1)
$$
$a$ and $n$ constants. Since these are relations taken from thermodynamics
one has a right to inquire about the temperature,  it was obtained from the ideal gas law
$$
p = \R \rho T.\eqno(2.2)
$$
Together, the two relations imply that $\rho/T^n$ is constant.
The fact that stellar matter seems to behave like an ideal gas was noted with
amazement by Eddington. But this approach, its apparent naivit\'e notwithstanding,   is
characterized by strong internal coherence; it  may serve as  a paradigm
for parallell developments.

In the first place, since the temperature is relevant to understanding the phenomena,
it is not simply a question of the hydrodynamic relation between $p$ and $\rho$,
one needs both of the   relations that  characterize a simple fluid
or, alternatively, an expression for a thermodynamic potential. In the case of an ideal gas of particles with adiabatic index $n$,  the free energy density is expressed in terms of its natural variables as
$$
f(T,\rho) = \R \rho T \ln{\rho\over T^n}.\eqno(2.3)
$$
The entropy  density is defined  as
$$
s = -{\p f\over \p T} =- \R \rho(\ln{\rho\over T^n}-n).\eqno(2.4)
$$
That is:  using the polytropic relation (2.1) with $a$ constant implies that the specific entropy density $s/\rho$ is a constant throughout.

The thermodynamic pressure and the chemical potential are  defined by
$$
p(\mu,T) = \rho{\p f\over \p \rho}\Big|_T-f(\rho,T), ~~ \mu := {\p f(\rho,T)\over \p \rho}.
$$
The pressure is thus a thermodynamic potential   defined by a Legendre
transformation; the natural variables are $T$ and the chemical potential $\mu$, with
$$
{\p p\over \p T}\Big|_\mu = s,~~ {\p p\over \p\mu}\Big|_T = \rho.
$$ 
In the case of an ideal gas this gives the ideal gas law 
$$
p= \R \rho T,~~ \mu = \R T\exp(\ln{\rho\over T^n}+1).
$$
and
$$
p(T,\mu) = \R T^{n+1} \exp({\mu\over \R T^n}-1).
$$

The Gibbsean action principle (minimum energy) leads to the following equation of motion,
$$
{\p \over  \p \rho}[\phi\rho + f(T,\rho)] = {\rm constant}.\eqno(2.5)
$$
where $\phi$ is the gravitational potential. By making use of the relations (2.1) and (2.2),
or more generally by  postulating that the specific entropy density is uniform, one can
derive the famous hydrostatic condition
$$
\rho\,{\rm grad}  \,\phi + {\rm grad} \,p =0.
$$
The \underbar{necessary} condition on the entropy reflects the fact that no energy source is being taken into account.  The problem of incorporating the effect of radiation   is solved in the Eddington approximation by adding the Stefan-Boltzmann term  $\hat a T^4$ to the (free) energy,
which generates the addition of  $(\hat a/3)T^4$ to the pressure. But we shall leave that refinement aside in our discussion.

  The usual system of equations that is used to describe a stellar atmosphere  consists of 
a relation between $p$ and $\rho$, with the continuity equation and the hydrostatic condition and, eventually, Einstein's equations. The relation 
between $p$ and $\rho$ must be interpreted as the result of the elimination  of $T$ in favor of the entropy,   which is usually taken to be uniform.  If the entropy is not specified   the eventual  success
of an equation of state, as usually understood to mean a relation between $p$
and $\rho$, can have many different  thermodynamical interpretations.

For  example, given any expression $p(\rho)$, one is free to postulate that the entropy is zero, implying that the free energy is independent of the temperature.
In this paper we shall assume that the entropy of dark matter is zero; that goes a long way towards explaining why it is difficult to establish any communication with
dark matter, other than via the intervention of gravity; it has no information to give us. 
But we  stress that the equation of state that we shall  use may  have  other interpretations; interpretations that can be distinguished when we gain
some independent knowledge about the temperature or the entropy of dark matter.

The search for an understanding of very dense stars led Chandrasekhar (1935),
following a suggestion by Stoner,  to an equation of state for partly or fully degenerate
Fermi gas. The relation between $p$ and $\rho$ is given in parametric form, the 
following expressions taken from Landau and Lifschitz (1959), 
$$
p = {2\over 3m^2}{(kT)^{5/2}\over m^2}A\int_0^\infty{\zeta^{3/2}d\zeta\over \e^{\zeta-t} + 1},
$$
$$
\rho = (kT)^{3/2}A\int_0^\infty{\zeta^{1/2}d\zeta\over \e^{\zeta-t} + 1},~~ A = {\rm constant},
$$
with the real parameter $t$. Since the natural variables of $p$ are $T$ and the chemical potential $\mu$, we expect that $t$ must be related to $\mu$ and indeed
the identification
$$
t= mc^2/2\hbar T
$$
leads, after a rescaling of the integration variable, to the formulas
$$
p(\mu,T) = {2\over 3}B \int   {\zeta^{3/2} d\zeta\over \e^{{mc^2\over2\hbar T}(\zeta - \mu)}+1}
$$
$$
\rho(\mu,T) = B \int    {\zeta^{1/2} d\zeta\over \e^{{mc^2\over2\hbar T}(\zeta - \mu)}+1}
d\zeta  = {\p p \over \p \mu}\Big|_T.
$$
where $B$ is a constant. These expressions are in Chandrasekhar (1957) page 400.
The first relation is enough to define the thermodynamical system.

To use this equation of state in the context of stellar atmospheres more information is required. One could, for example, fix the temperature.\footnote * { See for example the discussion in  a footnote  in Oppenheimer and Volkov (1939).}
 More in line with the earlier work based on the ideal gas would be to fix the entropy density
$$
s = {\p p\over \p T}\Big|_\mu.
$$
Numerical calculations show that the entropy is in fact very low for temperatures that
are not extremely high; therefore,  fixing the temperature is roughly equivalent to setting the entropy to zero.

Let us examine this theory in  the degenerate limit that Chandrasekhar 
(1957, page 358) describes thus: ``a completely degenerate electron gas is one in which all the lowest quantum states are occupied". Taking the limit in this sense Chandrasekhar ends up with a hydrodynamical equation of state, also given parameterically, as follows\footnote * {Oppenenheimer and Volkov used a variant  of this equation of state in their
pioneering work on neutron stars.   We were not able to relate their parameter (``$t$")  to a chemical potential. }  

$$
p(T,\mu) =  {B\over 3}{\int_0}^{\theta_0}\sinh^4\theta\, d\theta,
$$
$$
\rho =   B\sinh^3\theta,~~ \cosh\theta_0 : = \mu. 
$$
This is the result of an approximation that involves replacing the exponential term
in the denominator by a cutoff at $\zeta = \mu$. What is important here is that
$$
{d p(\mu)\over d \mu} = \rho(\mu),
 $$
since it identifies the parameter $\mu=\cosh\theta$ as the chemical potential.
In this approximation the dependence of $p$  on $T$ is neglected;
so that the entropy is effectively zero. 
In the next section we shall propose an equation of state that bears a strong resemblance to these relations:
$$
\rho(\mu)=A\sinh^4\mu,~~ p(\mu) = A \int_0^\mu d\nu\,\sinh^4\nu\,\,d\nu.
$$
Here too,  $\mu$  is the chemical potential.

 \bb
 
\b\b\no{\bf 3. The equations of motion}

We shall calculate static, spherically symmetric  solutions of Einstein's equations,
$$
G_{\mu\nu} = {8\pi G\over c^2} T_{\mu\nu},~~ G = .7414\times 10^{-28} {cm\over g},
$$
with a metric of the form
$$
ds^2 = \e^\nu (cdt)^2 - \e^\lambda dr^2 - r^2d\Omega,~~
g_{00}= c^2 \e^{\nu(r)},~~ g_{rr} = -\e^{\lambda(r)},
$$
and a matter energy momentum tensor of the form
$$
T_{00} = \rho U_0U_0,~~ T_{rr} = pg_{rr}, 
\eqno(3.1)
$$
all other components zero. Besides Einstein's equation we invoke the hydrostatic
condition in integrated form\footnote {$^1$}{See Section 6.} 
$$
{c^2 \over 2}(\e^{-\nu} -1) = {d f\over d
 \rho}.\eqno(3.2)
$$
 
\b

 The reduced form of Einstein's equations given in the textbooks, beginning with that of Tolman (1934),  is \footnote {$^1$} {Tthe prime stands for differentiation with respect to $r$.}
$$\eqalign{&
\hskip1cmG_t^t =  -e^{-\lambda}\Big({-\lambda'\over r}+ {1\over r^2}\Big) +
{1\over r^2} =  8\pi G \Big(\e^{-\nu}\rho -p/c^2\Big),\cr 
&\hskip1cm {G_r}^r =  -{\rm e}^{-\lambda }\Big({\nu'\over r} + {1\over r^2}\Big)+
{1\over r^2} =  - 8\pi G   p/c^2.\cr
\cr}\eqno(3.3) 
$$
With the notation \footnote{$^2$}{The choice of the letter $m$ in the first expression is traditional, but unfortunate, in as much the locally observed mass 
defined in (1.1) is $2M(r)G = m(r) + u(r)$. (See below.)}
$$
H(r)=\e^{-\lambda} = 1 -{ m(r)\over r},~ K(r) = \e^{\nu +\lambda} = 1 -{u(r)\over r},
~~HK=\e^\nu = 1 - {\phi\over r};
$$
they are
$$\eqalign{&
H' = {1-H\over r} - 8\pi G r  \rho\big(\e^{-\nu} - p/c^2\rho\big),\cr
&
K' =8\pi G H^{-2}  r \rho.\cr}
$$
  
 To continue we need an expression for the free energy that will allow us to express
 the density and the pressure in terms of the fields, with the help of Eq.(3.2). To determine the free energy we shall work, provisionally, with the weak  field approximation. 
\ve

\no {\bf 4.  The equation of state} 

A weak field approximation will be used to determine an approximate equation of state. In this approximation we replace Eq.s (3.4) by
 $$
 m'(r) = \, r^2w \rho,  \,\eqno(4.1)
 $$
 $$
 K'(r) =  \,r\,w\rho.\eqno(4.2)
 $$
 The primes, as before, denote the derivative with respect to $r$.  The   newtonian potential is $-\phi/2$, where
 $$
 1-\e^{-\nu}\approx (1-H) + (1-K) = {m \over r} + {u \over r} = :\phi.
 $$
 Combining (3.1-2) we get
 $$
 m= -r^2 \phi',~~ w\rho=m'/r^2
 $$ 
  The full set of equations is thus
$$
{c^2\over 2}\phi=  {\p f\over \p \rho},~~~  -r^{-2}(r^2 \phi')'= w\rho.  \eqno(4.3)
$$

Something is known about  $\phi$, from observation of radial acceleration of orbiting stars. A family of  satellites moving in circular   orbits with radius $r$ in a  radial,
newtonian  potential  $V$ have  orbital speed $v$ given by $v^2 = rV'$.  Observation has 
revealed that there is a wide interval in which the speed is nearly constant, independent of the distance, which implies that, in this interval, the potential is
approximated by $V = -\phi/2=(k/2)\ln(r)$. We shall model the function $\phi(r)$, then 
calculate the equation of state.  In other words, when the distribution $\phi(r)$ is known from observation, then the last pair of equations provides a parametric representation of the relation between the free energy  density $f$ and the density $\rho$.
Finally  we shall use this equation of state in the exact, relativistic 
field equations. 
\bb
\ce {\bf Example} 

Taking
$$
\phi =k \ln{r + b \over r},~~ r = \e^x,
$$
we obtain the velocity distribution shown in  Fig.1  with $k=10^{-6}$,   $b = \e^{54}$.
 It is very nearly constant for $x<52$ and very nearly newtonian outside.
 In the flat region the value is
$$
 v/c= \sqrt{-{1\over 2} r\phi'(r)} \approx \sqrt{-k/2},~~ v=212\, km/sec.
 $$
 The value of the constant $k$ is thus determined by observation of the orbital velocities.
 
 In the region of the innermost orbiters $r  \approx r_0 =10^{16}$ and the local mass is
 $$
 2mG = r \phi = r _0 k \ln{r_0+b\over r_0}.
 $$
 With $b = \e^{54} = 2.83\times 10^{23}$ we obtain $2mG = 1.7\times 10^{11}cm$ which is close enough to the observed value $10^{12}cm$.

\vskip4cm

\parindent=1pc

\vskip-3cm
 
\epsfxsize2.5in
\centerline{\epsfbox{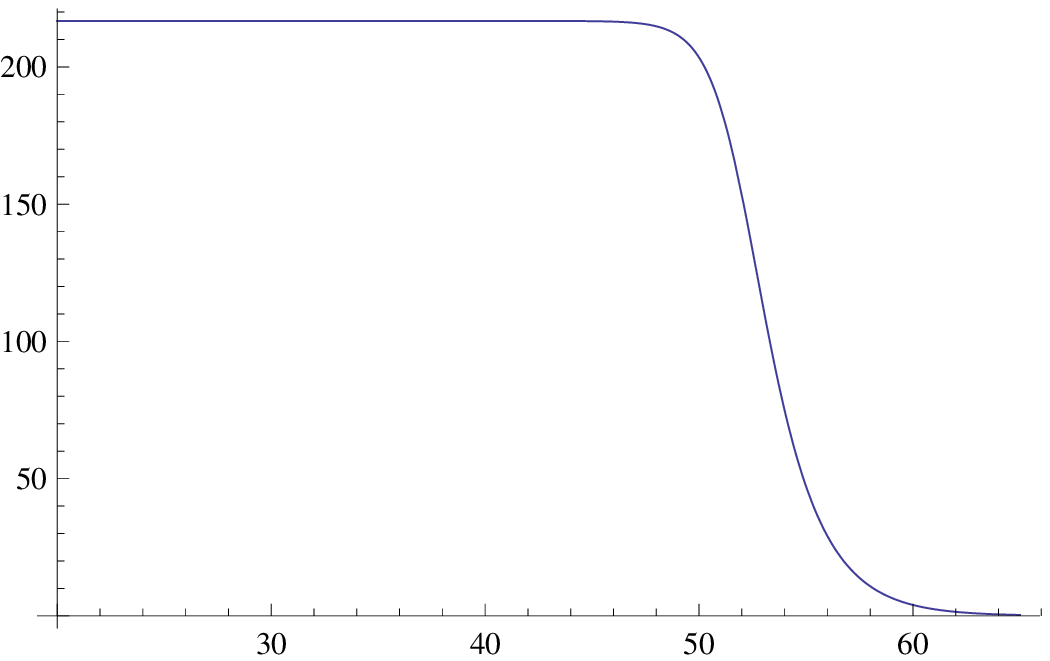}}
\vskip1mm
 
 Fig.1. The orbital velocity distribution  that  was used as a model of the observations in the Milky way, for $k=1.044\times 10^{-6}.$  The velocity is shown in km/sec. 
 \b\b

 From (3.4) we get
 $$
m(r)=-r^2\phi' = kb{ r\over r+b} ~~(= kb - {kb\over r} + ...,~~ r>b),
$$
and the density
$$
w\rho(r) = -\Delta\phi=-r^{-2}(r^2\phi')'= {k\over r^2(1+r/b)^2} .\eqno(4.4)
$$

 The parametric representation of the equation of state is thus, in this case,
$$
 {2\over c^2} {\p f\over \p \rho} = \phi(t) =k\ln{t + b \over t} ,~~w\rho(t) = {k\over t^2(1+t/b)^2}.
$$
Equivalently,
$$
\rho = {16\over b^2}{k\over w}\sinh^4{\psi}, ~~ \psi:= \phi/2k.\eqno(4.5)
$$
The free energy is obtained  by integrating the hydrostatic equation,
 $$
 {2\over c^2}{df \over d \rho} = \phi,~~ {2\over c^2}{df \over d \psi} = \phi{d\rho \over d \psi} =
  2k\Big({d\over d\psi}(\psi\rho) -   \rho\Big).
 $$
 Thus
 $$
 \hat f(\rho)={b^2w\over 16 c^2 k^2 
 }f(\rho) =\int \psi{d\over d \psi}\sinh^4\psi  \, d\psi =\psi \sinh^4\psi-\hat p.
 $$
The last term,
 $$
\hat p := \int\sinh^4\psi\, d\psi = {1\over 32}\Big( \sinh(4\psi)- 8\sinh(2\psi) + 12\psi\Big),
\eqno(4.6)
 $$
 is the pressure,
 $$
p= {16\over b^2}{k^2c^2\over w}\hat p  = \rho{\p\over \p \rho}f - f.\eqno(4.7)
 $$
 
\bb

{\bf Remark.} The function 
$$
\psi = {1\over 2}\ln(1+b/r)
$$
is an exact solution of the modified Emden equation (4.4)
$$
\Delta \psi + {8\over b^2} \sinh^4\psi = 0,~~ \Delta = r^{-2}{d\over dr} r^2 {d\over dr}.
$$
The original Emden equation has $\psi^n$ instead of  $\sinh^4\psi$; it has an exact solution in the case that $n = 5$ only.

 \bb

Returning to the equations (3.1-2) - General Relativity -  we now 
apply the equation of state in the form (4.5-6), fix the appropriate initial values,
$$
x =a=52,~~m(a)={ bk\over 2} = .2\times 10^{17},~~k= 1.044 \times10^{-6},~~K(a) = 1-k\ln2 +{k\over 2},~~\phi = k\ln 2
$$ 
and run the equations in Mathematica. The program runs from
 $ x = 52 (r=4\times 10^{22})$ in both directions, covering 36 orders of magnitude of the radius, correctly reproducing the exact solution (3.4) from $x = 0$ to $x = 83$ ($r=0$ to $r=10^{36}$cm).

 \bb

\no{\bf 5. Solutions of the relativistic equations of motion}

It is customary, whenever it is difficult to construct a physically reasonable model of a stellar object, to conclude that one must be dealing with a Black Hole, which in the astrophysical context, if taken literally, means a Schwarzschild Black Hole. It is a conjecture that is difficult to verify, since it is impossible to receive information from the horizon,
let alone from beyond. And then there is the very thorny issue of the creation of a Black Hole, since it is difficult to describe a reasonable scenario of matter ``falling into" it.  But the main difficulty is that a Black Hole is a property of empty space.

All ``theorems" that claim to limit the size of any kind of stellar object rely on some
assumption about the equation of state.  Early work was entirely based on polytropes that are notorious for unphysical properties. Already the work of Emden,
 100 years ago,  show that no choice of the polytropic index is satisfactory, either because the mass is infinite or because the density (or the pressure) turns negative at a finite distance from the center.  The impractability of carrying out a thorough program of numerical calculations was  a limiting factor 50 years ago as is
exemplified by the well known paper by Oppenheimer and Volkov (1939), the first 
attempt to use a much more sophisticated equation of state.  And even today it is difficult to determine the intial conditions that characterize the very elusive global solutions. This paper adopts an equation of state that resembles that of Oppenheimer and Volkov (1939), especially appropriate for Dark Matter. A powerful computer, and a novel strategy for discovering the elusive initial values that are needed  for numerical calculations allow us to do calculations that have not been feasible in the past. What follows is the result of a study of thousands of solutions.

It is not  difficult to find equations of state that give rise to regular, generally relativistic gas spheres of very high mass. In this paper we are making use of the fact that  the equation of state of Dark Matter is  not known from other evidence. The result  is of general interest, for
if we can visualize a reasonable object consisting mostly of Dark Matter then there should be no difficulty in imagining that ordinary matter can stick to it, or fall into it.

  Imagine traveling towards the center of an object of a certain mass $M$. This mass is determined by measuring the metric at great distances from the center of the object, where to a good approximation the density of matter is zero. The mass  $M$
determines the Schwarzschild radius. In the case of a galaxy like our own this radius is  $10^{16}-10^{17}$cm.  But  here we are  closer to the center
than the greater part of the mass.
Measuring the metric at this neighbprhood we find that the Schwarzschild radius has shrunk to $10^{11}$ cm.  And if we should have any prospect of finding a horizon at that place then  there must be little or  no matter in the interval.
  
Given a value of the total mass, to encounter a phenomenon that resembles a horizon we must arrange for most of the mass to be squeezed into a configuration that has a very low density everywhere except for a small region very close to the center. A sequence of configurations, of increasing concentration near the center, is provided by our equation of state:  by keeping the total mass  $bk/2$ fixed and gradually decreasing the effective radius ($b$).  Once  we have a family of models of stars with a fixed total mass, indexed by a parameter that controls the effective size, then we have a tool for trying to understand if and  how a star can  acquire a horizon. This sequence is described in detail in Subsection 5a.

Every effort to understand stellar evolution describes
evolution, a relatively slow process, as a progression of adiabatic equilibria,
configurations that are equilibria in the short term.  Without describing the dynamics of evolution  we propose to
form
 an evolutionary sequence from our family of static models. The principal  constraint is that each adiabatic equilibrium configuration must be a stationary solution of Einstein's equations. Now the successive equilibria have to
be plausibly linked by evolution, and that brings us to the important question of mass and conservation laws.

Although we shall not attempt to describe the evolutionary process, it is possible to speculate about the behavior of the principal parameters of the evolving stellar object.  Thus it seems at first quite natural that the total mass remain unchanged. An argument
can be made to show that the mass is likely to diminish, and with some cost
of plausibility one can imagine conditions under which it may be increasing.
The neutral point   among these contrasting scenarios is the case that the mass remains fixed
during the evolution; this will serve as a natural bench mark  and it is natural to investigate this possibility first.

But why focus on the mass? And what exactly is the mass? If the change of some quantity is characteristic of   evolution, then it must be  conserved  by the adiabatic dynamics.

Tolman's phenomenological approach (Tolman 1934) incorporates a density that is tentatively interpreted either as a density of mass or a density of particles, but this density is not conserved. Indeed, the theory abandons one of the two principal features of
hydrodynamics.  The total mass, as defined asymptotically by the limiting Schwarzschild metric can be expressed, in Tolman's theory as
$$
M =4\pi \int_0^\infty m(r) r^2dr,
$$ 
where the function $m(r)$ is defined as
$$
{m(r) \over r}={1
\over 2} ({1\over g_{rr}}-1),
$$
in terms of the radial component of the metric. The identification of this function as  density is common, although, as remarked by Kippenhahn and Weigert  (1990)
and others, it is hazardous, since its transformation properties are not correct.  

In this paper we are using  an approach to Relativistic Thermodynamics that, at first sight, and in a restricted context, is indistinguishable from Tolman's theory.  It is based on Gibb's  thermodynamical action principle  and incorporates hydrodynamics with a conserved density.
In the newtonian approximation it is possible to interpret this as a mass density, but in the relativistic context we have adopted another use of the word ``mass",
\footnote {$^3$} {Introduction, Eq. (1.1).} and there
is no compelling reason to relate the two concepts. We shall abandon the concept of ``mass density" and we shall refer to the conserved density 
$\rho$ as the particle number density.  In the case of the model galaxy, with
parameters $b = 52$ (radius $R=3.8\times\e^{22}$ cm) and total mass $bk/2=2\times 
10^{16}$ cm,  we find
$$
N:= 2Gc^2\int \rho ~g^{00} \sqrt{-g}r^2dr~ \rho= 1.999\times 10^{16 
} cm.
$$
The close agreement of this number  with the gravitational mass (when $\ln b = 52$)  is due to the fact that the newtonian approximation is valid in the greater part of the galaxy. 
  
  As we reduce the dimension (the value of the parameter $b$), we now have to
reduce $k$ as well, to keep constant the number of particles. The sequence that 
results from this procedure is described in Subsection 5b.  

 \b

\ce{\bf 5a. A sequence with fixed Schwarzschild  mass}

We need the full dynamics of General Relativity; the equations were summarized in Section 2. To begin, we solve the non relativistic problem and note the initial values at $x = 52$. This gives us initial values to be used in Einstein's equations,
otherwise very difficult to determine.  No assumptions are made initially concerning the behavior of the solution near the center, for all experimental information pertains to the outer region. 
Once it has been established  that the computation can be extended to $r=0$
we  can start at either end.
 
To reach the center  
it is necessary to fine tune the initial values to more than 10 significant digits, although the adjustment is tiny, never more than 1 percent and in most cases less than one part in 10 000. With a fair amount of labor
we succeeded in running from  $r=\e^{52}cm$  or more to within 1 cm of the center
and outward to about $\e^{70}$.

We started with the parameters that fit the Milky Way and fixed the mass $kb/2$.
With initial value $\ln  b = 52 (R=3.8\times 10^{22})$ we reduced this parameter by  steps to $\ln b = 40 (R= 2.4\times 10^{17})
$, 
reducing
 the radius of the model galaxy by 5 orders of magnitude. Here is what we found.

In an outer region the newtonian approximation is surprisingly accurate and the 
parameters $b$ and $k$ determine,  via the sinh-Emden equation, an 
essentially unique solution. As the mass is compressed, by reduction of the extent (radius) $b$ with the mass $bk/2$ kept fixed, this region grows. That is, the 
density of mass (particles) in the outer
region, the region that is accessible to observation, becomes very small, and this
region grows.

\vskip4cm

\parindent=1pc

\vskip-3cm
 
\epsfxsize2.5in
\centerline{\epsfbox{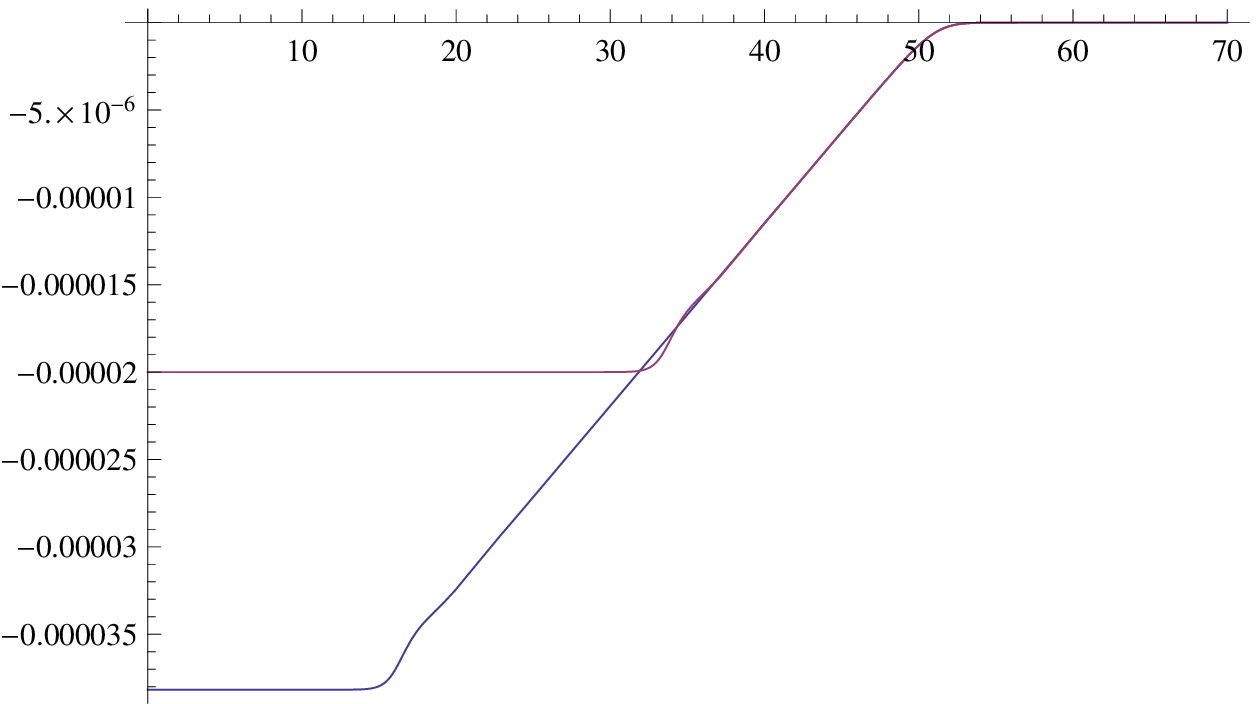}}
\vskip1mm
 
 Fig.2.  The metric function $K-1$, obtained by integrating from the center outwards,
 with different initial values. From $x=30$ the solutions are indistinguishable.
 In this example $\ln b = 52$ and $m(0) = 0$.
\b
In an inner, complimentary region there is a great disparity between solutions. Fig. 2 
shows a typical example of the confluence of two solutions that start off very different near the center. Conversely, when a solution is extended inwards 
its course cannot be predicted, for it depends on extremely minute fine tuning of the initial values of $m$ and  $K$.  
 
A behavior hinting at the development of a Black Hole would be a rise in the value of the potential $\phi$ near the Schwarzschild radius.  \underbar{ No tendency
towards such development} \underbar{ was observed. }  Fig. 3 illustrates the dramatic rise  in the values of the potential, by a factor of 500,  as $\ln b$ is reduced from 52 to 46. 
\vskip4cm

\parindent=1pc

\vskip-3cm
 
\epsfxsize2.5in
\centerline{\epsfbox{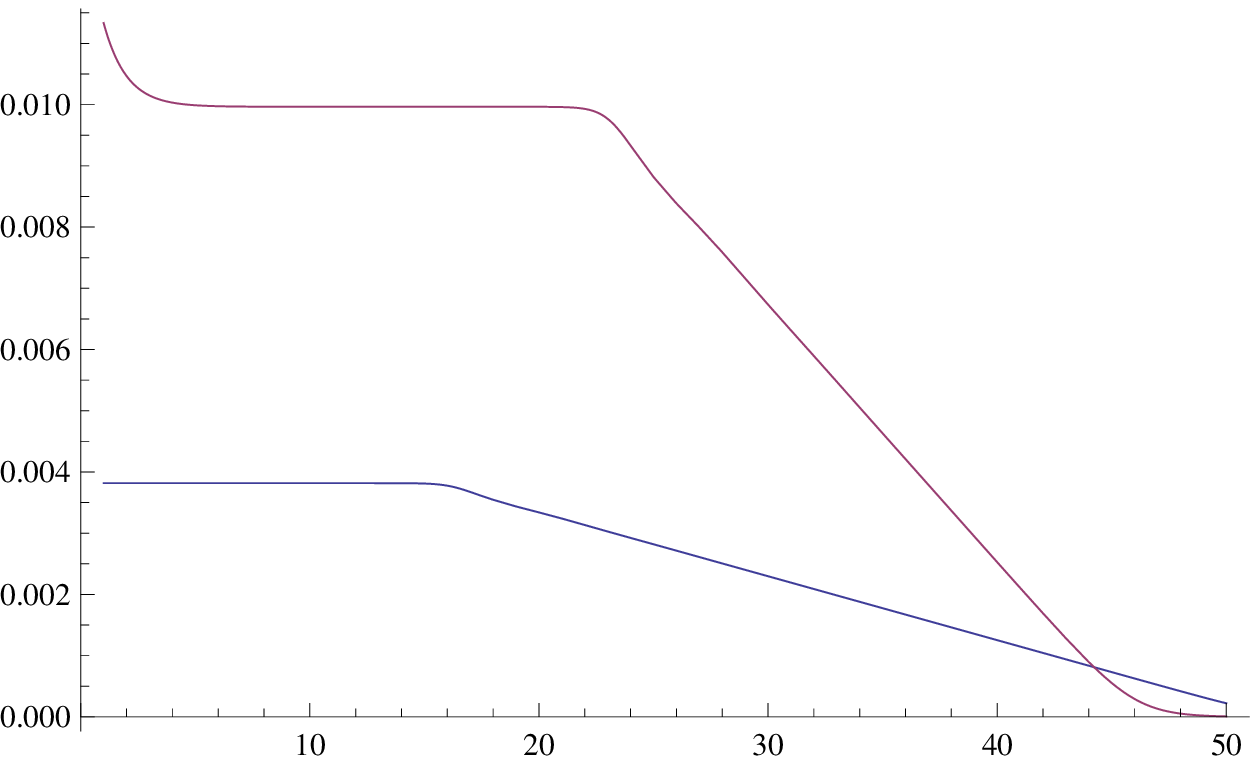}}
\vskip1mm

Fig.3. The metric function $\phi$ (the gravitational potential) in the case that $\ln b = 46$  (upper curve) compared with $100 \phi$ in the case that $\ln b = 52$ (lower curve) .
\b
All integration was first done from the outside in, then the final values of the metric functions at $x = 0$ were used as initial values for the reverse run along an identical track.  

 The orbital velocity changes in an interesting manner, as illustrated in Fig.4 in the case $\ln b = 46 (R= 10^{20})$. The high velocities that were introduced by the choice of equation of state last only as far as $x =20 (R = 5\times 10^8)$, but  return to high values very near the center.  This behavior of the orbital velocity suggests that the very high velocities of the inner orbiters of the Milky Way may be attributed to General Relativity, in a model with a more refined equation of state. 
 
 \vskip4cm

\parindent=1pc

\vskip-3cm
 
\epsfxsize2.5in
\centerline{\epsfbox{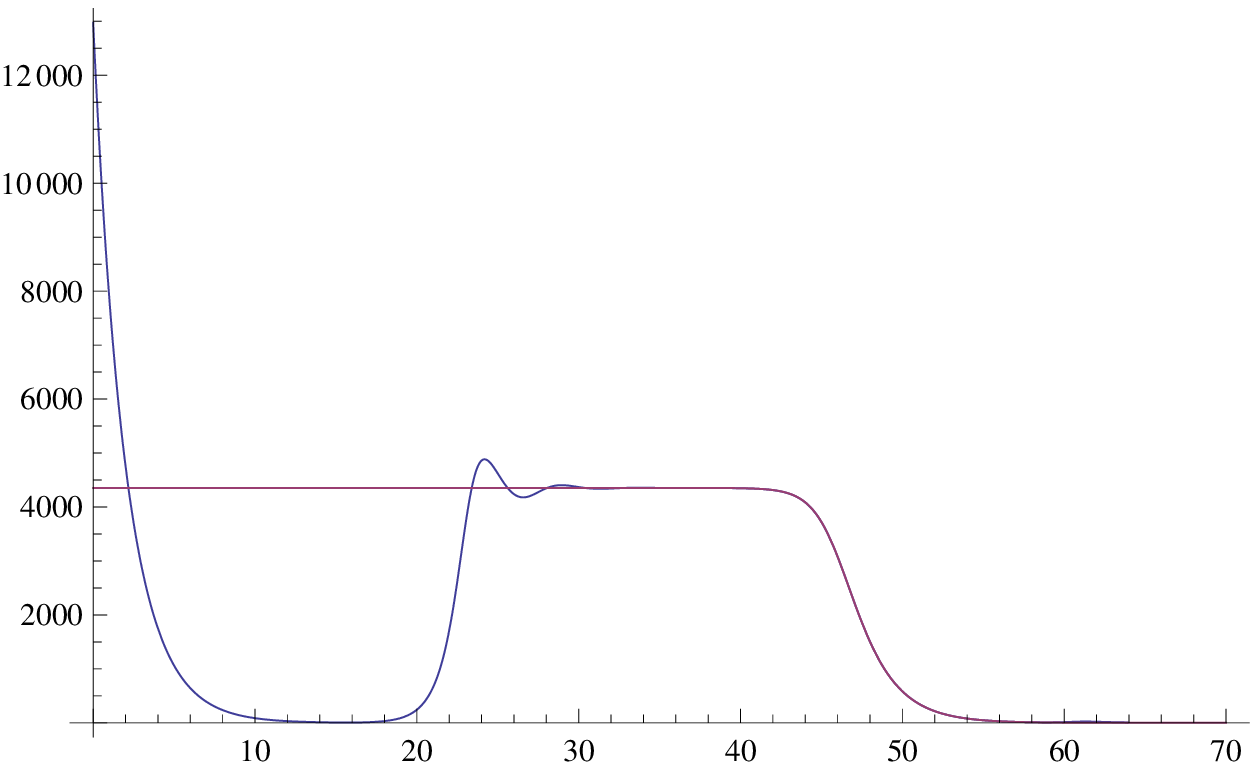}}
\vskip1mm

 Fig.4. The orbital velocity in the case $b=46$. The horizontal line is the result
 of the newtonian model. The drop at around $x = 20$ is characteristic, the increase
very close to the center is not - it depends on the intitial values and it may be spurious.

\b

 Finally there is the important question of the density. Fig. 5 shows the 
 profile in a   log-log presentation (natural logarithms). In the innermost region the density is   high, about $10^7 g/cm^3$.
 But at $x=30$, which is far closer than the inner orbiters, it is $5\times 10^{-6}$. 
 
 As we squeeze the mass into a smaller region the density near the center grows,
we obtain density profiles that  vary greatly with the boundary conditions at the center. The densities can get very high.

\vskip4cm

\parindent=1pc

\vskip-3cm
 
\epsfxsize2.5in
\centerline{\epsfbox{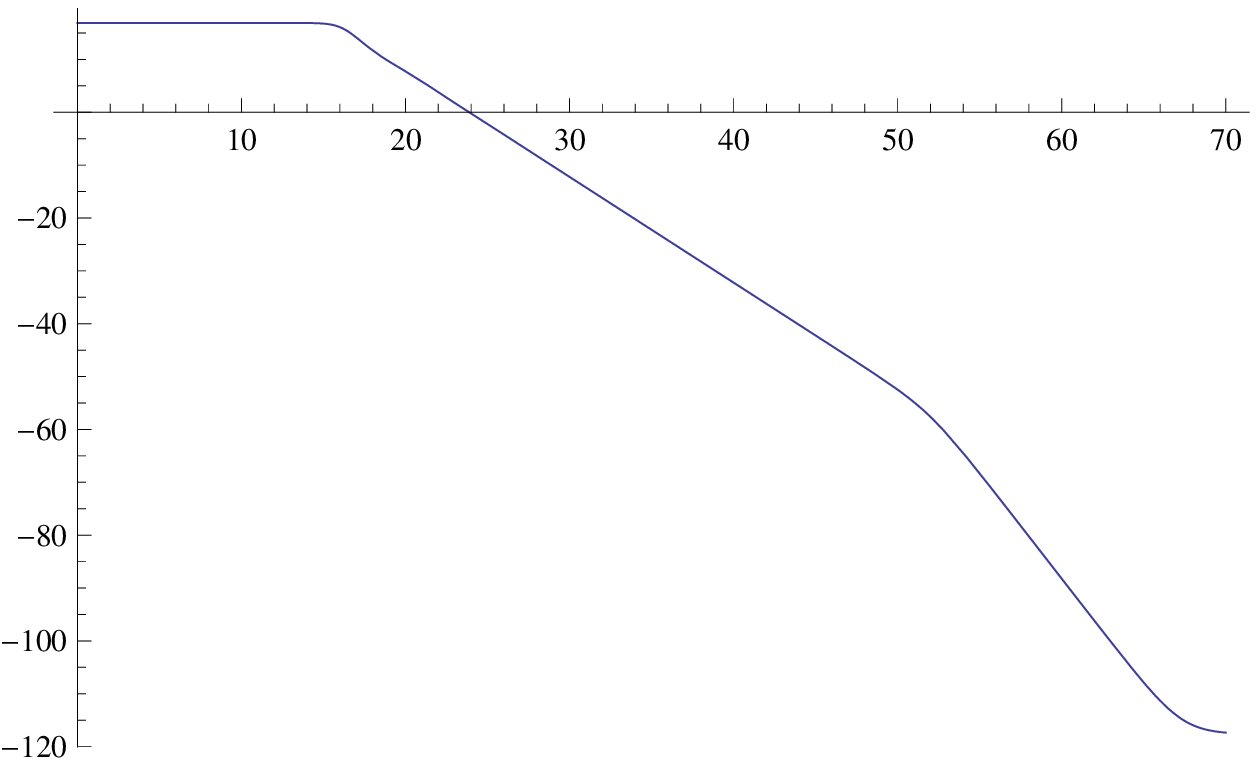}}
\vskip1mm

Fig.5. A log/log plot of the density profile in the galactic model, $\ln b = 52$.
\b

The first column of the following  table give some values of $N$ in the case of the galactic model with $b = 52$ and Schwarzshild mass $bk/2 = 2\times 10^{16} cm$. The integration runs from $x=0 ~ $to$ ~x=X$. Each of the other columns show the result of reducing the size, to $R = \e^{50}, \e^{48}, \e^{46}, \e^{44} cm$, respectively.
\b
\settabs \+      & 
~~~~~ &52~~~~~~~~~
~~~& ~~50~~~~~~~~~~~&
48~~~~~~~~~~~~~&   $3.5\times 10^{16}$~~~&  ~~~~44~~~~~~~~&4$3.5\times 10^{16}$~&$3.5\times 10^{16}$~ \cr

\+&X&$\ln b=52$~~~~&$\ln b=50$&$\ln b=48$&$\ln b=46$&$\ln b=44$&$\ln b=42$&$\ln b=40$\cr

 \+&52&$2\times 10^{16}$&$3.5\times 10^{16}$&$3.9\times10^{16}$
 &$4.0\times 10^{16}$&$4\times 10^{16}$&$8.3\times 10^{17}$&$9.3\times 10^{17}$\cr

\+&48&$7.2\times 10^{14}$&$4.8\times 10^{15} $&$ 2.0 \times 10^{16}$
&$3.5\times 10^{16}$&$3.9\times 10^{16}$&$4.5\times 10^{16}$&$ 4.4 \times 10^{16}$&\cr 

\+&44&$1.3\times 10^{13}$&$9.9\times 10^{13} $&$ 7.2\times 10^{14}$
&$4.8\times 10^{15}$&$2.0\times 10^{16}$&$3.2\times 10^{16}$&$
3.4\times 10^{16}$\cr 

\+&40&$2.5\times 10^{11}$&$1.8\times 10^{12} $&$ 1.3\times 10^{13}$
&$9.8\times 10^{13}$&$7.1\times 10^{14}$&$4.4\times 10^{15}$&$4.1\times 10^{15}$\cr 

\b

 All our solutions were checked by integrating from the center outwards as well as inward from a large distance.   The last column gives an indication of why further reduction of size (beyond $\ln b = 44$) is not possible.   All attempts to reduce $b$ below $\e^{40}, (R=2\times 10^{17})$ were unfruitful.

The main conclusion is that (this model of) the Milky Way can contract  by at least
five orders of magnitude of linear dimension  without the need to develop a horizon.

\bb

\ce{\bf 5b. An evolutionary sequence}

With the same starting point, at $b = 52$ and $bk = 4\times 10^{16}$, we now 
reduce both parameters while holding the number  $N$ fixed at $x = \ln r = 52, 
N(52) =  2.5\times 10^{17} cm$.  The second table shows a significant departure from the
first one. The ``mass", in cm, is the $2G$ times the Schwarzschild or asymptotic mass.
 \b
\settabs \+ ~     & 1~~~~~  & 20~~~ ~ ~~~~~~~& 100~~~~~~~~~~~~&
1000~~~~~~~~~~~& 10000~~~~~~~~~&~~~~~~~~~~~~~~~~  &
100000~~~~~& \cr

\+&X&$\ln b=52$~~~~&$\ln b=50$&$\ln b=48$&$\ln b=46$&$\ln b=44$&$\ln b = 42$ \cr

 \+&52&$2.0\times 10^{16}$&$2\times 10^{16}$&$2\times10^{16}$
 &$2\times 10^{16}$&$2\times 10^{16}$&$2\times 10^{16}$
 \cr

\+&48&$7.2\times 10^{14}$&$2.7\times 10^{15} $&$ 1.0 \times 10^{16}$
&$1.8\times 10^{16}$&$1.9\times 10^{16}$&$8.5\times 10^{15}$\cr 

\+&44&$1.3\times 10^{13}$&$5.7\times 10^{13} $&$ 3.7\times 10^{14}$
&$2.4\times 10^{15}$&$1.0\times 10^{16}$&$7.4\times 10^{15}$\cr 

\+&40&$2.5\times 10^{11}$&$1.0\times 10^{12} $&$ 6.8\times 10^{13}$
&$4.9\times 10^{13}$&$3.5\times 10^{14}$&$9.8\times 10^{14
}$\cr 
\vskip.5cm
\+&Mass & $2\times 10^{16}$&$1.15\times 10^{16}$&$1.01\times 10^{16}$&$1.0\times 10^{16}$&$1.0
\times 10^{16}$&$ 4.2\times 10^{15}$\cr
 \b
 The most remarkable feature is that, if the   current  conservation law is maintained by the evolutionary dynamics, then the mass, the observed Schwarzschild or
asymptotic mass of the galaxy,  will be reduced by a factor of 2  as the size of the galaxy is reduced by about 1 order of magnitude. It is not sure that the number of particles is preserved during evolution, but we do not expect it to increase; the estimate of a factor of  2  is thus a lower limit.

\bb
\no{\bf 6. Conclusions}
 
In this paper we have sought to determine a principal characteristic of 
 the distribution of dark matter that is believed to be largely responsible for the observed distribution of stellar velocities within our galaxy.  

An equation of state in hydrodynamics is a relation between density and pressure. The contribution of the pressure to EinsteinÕs equation is not as important as the role that is played by the pressure in the hydrostatic equation. In our approach the equation of state is expressed by the free energy density.  It is completely phenomenological, extracted from experimental data.
It is very reminiscent of the degenerate fermion equation of state proposed by Chandrasekhar in another context. Interpretation of the data thus suggests a similar
physical model for Dark Matter.  It is in the nature of the problem that the basic physics  is not known, it must be discovered by interpretation of the experiments.

Ultimately, the model of our galaxy must be improved by including the contribution of visible matter.  \footnote* {The fact that there is no interaction between dark and visible matter will make  this very straightforward;  for in this case  the free energy density is additive. Note that it is essential to recognize the roles of two quite different density fields. In this paper the contribution of the visible component is not taken into account.}

 The shape of the dark matter distribution is not known, but to make a start the  overall structure was studied in terms of an idealized, continuous, spherically symmetric distribution that makes  an important  contribution to the source of the gravitational field.  We used the observed data in the outer region to determine an equation of state. We assumed that the same type of dark matter is dominant in the central region, but this is inconsistent with   observations in a small region near the galactic center, generally associated with the compact radio source Sgr A*. 
 The stellar velocities observed in this region do not conform to the predictions of our model in its present form.      We interpret this as evidence
 for a second type of dark matter in the central region, or as the result of a phase change taking place at high pressures.
 
 In a subsequent paper we hope to report on an attempt to account for all the data by a two-component model of Dark Matter. In addition we try to find an efficient
way to include the visible component with its characteristic, disk-like shape.

\ve

 \no{\bf Acknowledgements}

We are grateful to   Paul Frampton, Philip Mannheim,  and John Moffat   for stimulating conversations.
 
\ve

\no{\bf References}

\no Ahmed, Z.  et al., (CDMS Collaboration), Phys. Rev. Lett., 102,011301 (2009)

\no Bilic, N.,  Munyaneza,  F. and Viollier, R.,   Phys. Rev. D, 59, 024003,(1998)

\no Chandrasekhar, S.,   Month.Not. R.A.S. {\bf 95} 222 (1935).  
 
\no Chandrasekhar, S., {\it An introduction to stellar structure}, U. Chicago press 1938.

\no Emden,  J. R., {\it Gaskugeln}, Teubner 1907.

\no Eddington, A.S.,    {\it The internal constitution of stars}, Dover. N.Y., 1959.

\no Fetter, A.L. and  and Walecka, J.D., {\it Theoretical mechanics of particles and continua},  

McGraw-Hill, N.Y. 1980.


\no Fr\o nsdal,  C.  ``Ideal stars and General Relativity", Gen.Rel.Grav. {\bf 39} 1971-2000 (2007).

\no Genzel,  R.,   Eisenhauer,  F. and  Gillessen, S., Revs. Mod. Phys., 82, 4, (2010)
 
 \no Ghez, A. M., Salim, S., Weinberg, N. N., Lu, J. R., Do, T., Dunn, J. K., Matthews, K., 

Morris,
M., Yelda, S., Becklin, E. E., Kremenek, T., Milosavljevic, M., Naiman, J., 

 ÒMeasuring
Distance and Properties of the Milky WayÕs Central Supermassive Black 
 
 Hole with Stellar
Orbits,Ó ApJ, 689, 1044 (2008).
 
\no Gibbs, J.W., ``On the equilibrium of heterogeneous substances",

Trans.Conn.Acad.  108-248 (1878).

\no Goobar, A.,  Hannestad, S.,  Mortsel, E.  and Huitzu, T J.,  Cosm. Astrop. Phys. 06 (2006).

\no Hartle, J.B. ``Bounds on the mass and moment of inertia of non-rotating neutron stars", 

 Physics Reports {\bf 46} 201-247 (1978).

\no Kippenhahn,  R. and Weigert, A., {\it  Stellar structure and evolution}, Springer-Verlag  1990.

\no Landau, L. D. and Lifshitz, E. M., {\it Statistical physics}, Pergamon Press 1958.	 


\no Munyaneza,  F., Tsiklauri, D. and Viollier,  Astrophys. J., 509, L105, (1998).

\no Oppenheimer, J.R. and Volkov, G.M., ``On massive neutron stars", 

Physical Review {\bf 55} 374-381 1939).

\no Stoner, E.C.,  ``The minimum pressure of a degenerate electron gas",  

Monthly Notices R.A.S., {\bf 92} 651-661 (1932).

\no Tolman,  R.C., {\it Thermodynamics and Cosmology}, Clarendon, Oxford 1934.  

\no Viollier,  R.D.,Trautmann,D, and Tupper,  DPhys. Lett. B, 306, (1993).

   \end
 
  The equations become more complicated but they can still be solved analytically. But this cannot be supported by observations. 
At this time there is no experimental information about the center of the Great Attractor; 
consequently it is not important to strain the model by extending the integration
much further in that direction. In particular, the integral of the function $m$, 
from the center outwards, contains no  information that is useful at this time.

\b

Other numerical results are as follows. The density is positive; there is a characteristic bump in the density 
profile  - see Fig.3b - where the density reaches the highest value, $ \rho =  100
g/cm^3$ at $x = 21.6$,
$$
\rho_{\rm max} = 100g/cm^3 ~~{\rm at}~~ r =   2.4\times 10^{9}.
$$
This ``object" is comparable to our Sun, in size, mass and gravitational field strength.
  At the shortest distance observed for an orbiting satellite, 
$r = 2\times10^{15} ~ (x = 35.23)$, the density is about $ 1.3\times10^{-10}$. The
pressure has a similar profile - Fig.4, with a peak value of  $2.5\times 10^{16}$.  The  gravitational field $-\phi(x) = c^{-2}g_{00}-1$ also has a maximum  -see Fig.5, reaching a maximum value of $\phi = .00003 $ at the same point. This is about  ten times stronger than the gravitational potential at the surface of the Sun. 
 The appearance of such  shapes is very common when 
polytropic equations of state are used; they are relativistic features not seen in the weak field approximation. For some stars the maximum value
of $\phi$ can rise to get very close to the limiting value of unity, at which point a horizon would appear.

\vskip3.5cm

\parindent=1pc

\vskip-3cm
 
\epsfxsize.6\hsize
\centerline{\epsfbox{Fig.3.eps}}
\vskip-.1cm

Fig.3. A plot of $\ln\rho/ \ln  r$ against $ x = \ln r$.
 \vskip2cm
\vskip2cm

\epsfxsize.6\hsize
\centerline{\epsfbox{Fig.4a.eps}}
\vskip0cm

Fig.4a. The characteristic, inner  density profile plotted against $x = \ln r$, with the peak at $r = 9.67 \times 10^{12} cm$.
\ve
\vskip3.5cm

 \vskip-3.2cm
 
\epsfxsize.6\hsize
\centerline{\epsfbox{Fig.1b.eps}}
\vskip0cm

Fig.5a. The pressure profile in the inner region.

\vskip3.5cm

\vskip-3cm
 
\epsfxsize.6\hsize
\centerline{\epsfbox{Fig.1b.eps}}
\vskip0cm

Fig.6. The potential has a maximum at the same point. The density and pressure peaks are narrower because of the high value of the ``polytropic index".

\b
\b

\parindent=1pc
\vskip3cm
 
\vskip-3.cm
 
\epsfxsize.6\hsize
\centerline{\epsfbox{Fig.2b.eps}}

Fig. 7. The velocity distribution predicted by the relativistic equations of motion, in km/sec.   Compare Fig.1. 
 \end
 
 Modify or omit this paragraph.  If this  equation of state turns out to be applicable in other galaxies as well,  then this approach to the problem of
dark matter can be considered as an alternative to modified gravity; see for example  Delbourgo (2008) and Mannheim (2011).

From this we shall determine the unique equation of state that is required in order that Einstein's equations admit this idealized potential in the   weak field approximation.

rts this view.rts this view.

Using the same equation of state we now solve Einstein's equations numerically.     
 
  The values of $m(r)$  and $r\phi(r)$ at some chosen values of the local mass, are
  as follows:
\vskip2mm\hrule\vskip2mm

\settabs \+ ~     & 1~~~~~  ~   ~~~~& 20~~~~~ ~ ~~~& 100~~~~~~~~~~~~&
1000~~~~~~& 10000~~~~~~&~~~~~~~~~  &
100000~~&~~~~~~~~~~~~&~~~~~~\~~~~~~~&~~~~~~~~~~~~~& Ideal \cr

\+&$r$&x&m&$r\phi$ \cr

 \+&$10^{10}$&23.03&9896&3.11$\times 10^6$\cr

\+&$10^{12}$&27.63&$1.04\times 10^6$&$2.54\times 10^7$\cr 
 
 \+&$10^{13}$&29.93&$1.04\times 10^7$&$2.26\times 10^8$\cr

 \+ &$10^{14}$& 32.24&$1.05\times 10^8$&$2.07\times 10^9$\cr
 
\+&$10^{15}$&34.54&$1.04\times 10^9$&$1.82\times 10^{10}$\cr 

 \+&$10^{16}$&36.84&$1.04\times 10^{10}$&$1.58\times 10^{11}$\cr

\+ &$10^{18}$& 41.45&$1.05\times10^{12}$&$1.11\times 10^{13}$\cr 
 
\+&$10^{20}$&46.05&$1.04\times 10^{14}$&$6.20\times 10^{14}$\cr 

 \+& $10^{22}$&50.66&$8.30\times 10^{15}$&$1.65\times 10^{16}$\cr
 
  \+& $3.8\times 10^{22}$&52&$2\times 10^{16}$&$2.77\times 10^{16}$\cr

\vskip2mm\hrule\vskip2mm

It t is not $m(r)$ that should be interpreted as the local mass, but $r \phi(r)$,
since $\phi$ rather than $m/r$  is the newtonian potential.  Although $K(r)$ remains small, the difference between $m$ and $\phi$ is considerable even at large distances. And at the distance of the inner orbiters the number $\phi(r)$  is certainly the number that
has been determined by observation of the orbital velocities. At that point it is  greater by a factor of 15 than the function $m(r)$, though still too small.

Observation of the innermost satellites of the Milky Way suggests a local mass of about
$ 10^{12} $ (3 million solar masses) at  a distance of $10^{16}$ from the center.

Tom does not like this paragraph. The local mass predicted by the model at the distance of the inner orbiters  is less than what is observed, by about one order of magnitude.  If the discrepancy can be removed by refinements of the model, then we shall have a picture of the galactic
center that is very different from a Schwartzschild black hole. A good fit to the mass of the inner orbiter is obtained by taking $b=\e^{54}$. The shoulder at $x = 50$ in the velocity distribution in Fig. 7 then moves outward to $x = 52, r = 2.8 \times 10^{22}$,
which may be too high to fit the observed radius of the dark matter 
distribution of galaxy. These numbers must remain somewhat uncertain until the
visible matter in the galaxy is taken into account.

  Upgrading the equations, from the weak field approximation (3.3) to the  exact   equations  of motion (3.4) has very little effect at the large distances that are the focus of the present study. It is of  some interest, nevertheless, to extend the solutions
  towards the center, where relativistic effects are important, in order to explore the possibility of an approach to a black hole. Calculations initiated at the largest distance come up against the limitations of our computer programs  at a distance of about $x = 20$. We shall explain how we circumvented this difficulty and 
  were able to get results that seem to be reliable, all the way to the center.
  
  Starting at $x=52$, we first verified that the numerical solution of the newtonian field
  equations is in perfect agreement
  with the analytical solution down to   $x = 20$ and beyond. The solution obtained by using Enstein's equations show only very insignificant deviations from it.
   
    With  a new initial point, at  $x = 30$, $r \approx 10^{13} cm$, with initial values obtained by the preceeding  calculation, the computation  runs all the way from the center to $r= 10^{35}$,  far beyond the limits of the universe, exactly reproducing the exact solution beyond $x = 30$.
This allows us to determine a range of initial values of $K$ and $m$ at the center, of solutions that are well behaved to the greatest distances. It is found that,
if $m(0)$ is kept fixed, \footnote * {All the calculations that are reported here used the following boundary values: $a=2, b= \e^{52}, c = .4\times 10^{17},w=1.863\times 10^{-27},  m(a)= -.000318$} and $K(0)$
is restricted to a narrow interval between 1-.00004 and 1-.00009,  the solution is very sensitive to the precise value of $K(0)$ chosen, and that dramatic features develop as the lower end of the range is approached. Thus Fig.s 3a-7a show the development of a strong discontinuity in the first derivative of the metric function $K$, as $K(0)$ is varied as indicated in the figure captions. Fig.s 4b-7b give close ups for the same intitial values. Fig.s 4c-7c show a parallell accentuation of a narrow peak in the newtonian potential $\phi$ and Fig.s 4d-7d are close ups.

The limitation of $K(0)$ to such a small interval does not indicate the approach of a singularity near the center; it is due to the requirement that the computation reach
all the way to the outer edge of the mass distribution. A remarkable feature of these 
solutions is that the total mass, as measured by the value of either by $r\phi(r)$ or
$m(r)$ at $x = 52$, are virtually unchanged as $K(0)$ is varied over the range.

And yet the solutions vary greatly. The function $K(r)$,  for example, changes dramatically, as shown in Fig. 9.  The figure shows two solutions, with
$K(0)$ at the ends of the range, $K(0) = 1-.000042$ and 1-.000089. Both solutions
agree to an amazing degree at large distances, after behaving quite differently from the center out to $x= 30$. 

The change in density is even more dramatic, as shown in Fig.10. The
density grows gradually inwards to $x = 30$. In the 1-.000042 solution
it continues to grow as the center is approach. In the 1-.000089
solution there is a sudden freeze, at a very low density, until between $x = 2$
there is a sharp peak. 

The impression that one  gets from this is that of a sequence of solutions of Einstein's equations in a near-vacuum. Progressing through the sequence one finds
that the near-emptiness extends further inwards, while the metric becomes more
and more disccontinuous in a region that gets smaller and smaller. A traveler
could get the idea that he is approaching a small black hole, for the density of matter remains very small until he gets very close. Small, because most of the mass of the Galaxy is outside his position. The local mass, as measured by the local quasi Schwartzschild metric, gets smaller with the approach.

Tom does not like this inner structure. I should try to get closer to a black hole.
Tom does not find this structure. I Think they are Ok except very close in.

\vskip1
cm
 
\epsfxsize.6\hsize

Fig.4a. A plot of $K(x)-1$ against $ x = \ln r$, $K(0) =   1-.000042$.  
  
\vskip.5cm
 
\epsfxsize.6\hsize
\centerline{\epsfbox{Fig.2a.eps}}

Fig.5a. A plot of $K(x)-1$ against $ x = \ln r$, $K(0) =   1-.000052$.

\vskip.5cm
 
\epsfxsize.6\hsize
\centerline{\epsfbox{Fig.3a.eps}}

Fig.6a. A plot of $K(x)-1$ against $ x = \ln r$, $K(0) =   1-.000062$.

\vskip.5cm

\epsfxsize.6\hsize
\centerline{\epsfbox{Fig.4a.eps}}

Fig.7a. A plot of $K(x)-1$ against $ x = \ln r$, $K(0) = 1-.000072$.

\vskip.5cm
 
\epsfxsize.6\hsize
\centerline{\epsfbox{Fig.5a.eps}}

Fig.8a. A plot of $K(x)-1$ against $ x = \ln r$, $K(0) =   1-.000082$.

\vskip.5cm

\ve
 
Fig.s 4b-8b are closeups of the same functions. 

\vskip.5cm
 
\epsfxsize.6\hsize
\centerline{\epsfbox{Fig.1c.eps}}

Fig.4b. A plot of $K(x)-1$ against $ x = \ln r$, $K(0) =   1-.000042$.

\vskip.5cm
 
\epsfxsize.6\hsize
\centerline{\epsfbox{Fig.2c.eps}}

Fig.5b. A plot of $K(x)-1$ against $ x = \ln r$, $K(0) =   1-.000052$.

\vskip.5cm
 
\epsfxsize.6\hsize
\centerline{\epsfbox{Fig.3c.eps}}

Fig.6b. A plot of $K(x)$ against $ x = \ln r$, $K(0) =   1-.000062$.

\vskip.5cm
 
\epsfxsize.6\hsize
\centerline{\epsfbox{Fig.4c.eps}}

Fig.7b. A plot of $K(x)-1$ against $ x = \ln r$, $K(0) =   1-.000075$.

\vskip.5cm
 
\epsfxsize.6\hsize
\centerline{\epsfbox{Fig.5c.eps}}

Fig.8b. A plot of $K(x)-1$ against $ x = \ln r$, $K(0) =  1-.000089 $.

Fig.s 4c-8c show the functions $\phi$ for $K(0)= $.  A sharp maximum appears, 
  very near   the center at $x \approx 5, r = 150cm$.

\vskip.5cm
 
\epsfxsize.6\hsize
\centerline{\epsfbox{Fig.1b.eps}}

Fig.4c. A plot of $\phi(x)$ against $ x = \ln r$, $K(0) = 1-.000042  $.

\vskip.5cm
 
\epsfxsize.6\hsize
\centerline{\epsfbox{Fig.2b.eps}}

Fig.5c. A plot of $\phi(x)$ against $ x = \ln r$, $K(0) = 1-.000052  $.

\vskip.5cm
 
\epsfxsize.6\hsize
\centerline{\epsfbox{Fig.3b.eps}}

Fig.6c. A plot of $\phi(x)$ against $ x = \ln r$, $K(0) =   1-.000062$.

\vskip.5cm
 
\epsfxsize.6\hsize
\centerline{\epsfbox{Fig.4b.eps}}

Fig.7c. A plot of $\phi\phi(x)$ against $ x = \ln r$, $K(0) =1-.000075$.

\vskip.5cm
 
\epsfxsize.6\hsize
\centerline{\epsfbox{Fig.5b.eps}}

Fig.8c. A plot of $\phi(x)$ against $ x = \ln r$, $K(0) =  1-.000089 $.

\b

 Fig.s 4d-8d are closeups of the function $\phi$.

\vskip.5cm
 
\epsfxsize.6\hsize
\centerline{\epsfbox{Fig.1d.eps}}

Fig.4d. A plot of $\phi(x)$ against $ x = \ln r$, $K(0) =   1-.000042$.

\vskip.5cm
 
\epsfxsize.6\hsize
\centerline{\epsfbox{Fig.2d.eps}}

Fig.5d. A plot of $\phi(x)$ against $ x = \ln r$, $K(0) =  1-.000052 $.

\vskip.5cm
 
\epsfxsize.6\hsize
\centerline{\epsfbox{Fig.3d.eps}}

Fig.6d. A plot of $\phi(x)$ against $ x = \ln r$, $K(0) =  1-.000062 $.

\vskip.5cm
 
\epsfxsize.6\hsize
\centerline{\epsfbox{Fig.4d.eps}}

Fig.7d. A plot of $\phi(x)$ against $ x = \ln r$, $K(0) =   1-.000075$.

\vskip.5cm
 
\epsfxsize.6\hsize
\centerline{\epsfbox{Fig.5d.eps}}

Fig.8d. A plot of $\phi(x)$ against $ x = \ln r$, $K(0) =1-.000089$.

\bb

\parindent=1pc
\vskip3cm

\epsfxsize.6\hsize
\centerline{\epsfbox{Fig.9.eps}}

Fig.9. The function $K(r)$ for the two extreme solutions.
\vskip1cm
\parindent=1pc
\vskip3cm
 
\vskip-3.cm
 
\epsfxsize.6\hsize
\centerline{\epsfbox{Fig.10.eps}}

Fig.10. The logarithm of the density for the two extreme solutions.
\vskip1cm

\parindent=1pc
\vskip3cm
 
\vskip-3.cm
 
\epsfxsize.6\hsize
\centerline{\epsfbox{Fig.11.eps}}
 
 Fig.10. Close up of logarithm of the density for the two extreme solutions.

\ve
\no{\bf V. The nature of dark matter}

The equation of state was obtained in parametric form, Eq.s (3.5) and (3.6),
$$
  \rho =C \sinh^4\psi,~~p = {kc^2\over 32}C(\sinh4\psi - 8 \sinh 2\psi + 12\psi ),\eqno(5.1)
$$
with
$ 
C = 16 k/b^2 \omega.
$
It bears a remarkable   similarity to an equation first proposed by Stoner(1932) and  used by Chandrasekhar (1935),\footnote {$^3$}{It was `corrected'  and used by Oppenheimer and Volkov  (1939)  in their study of  neutron stars.
The original  version, quoted here,  is in Landau and Lifshitz (1958) page 168.} 
$$
\rho = C_1\sinh^3t,~~ p =C_2 \big(\sinh 4t - 8 \sinh 2t + 12 t \big).
$$
The basis for this formula is a model of fermions in a collapsed state, the Femi sea being filled up to $q/m = \sinh\psi$.

The similarity, if not regarded as a coincidence, suggests that dark matter  may be a cloud of ``ice crystals" that consist of fermions in a highly reduced state.
The total absence of interactions, and of photons, is a premise of Chandrasekhar's work.
  A slightly different model reproduces our equations (5.1) exactly,
but the fermions need to have an additional degree of freedom; for example, an extra dimension of momentum space. With $q = \sinh\psi, E = \cosh\psi$, integrating over the 3-sphere with radius $q$,\footnote{$^4$}{  The factor $E$ in the first integral arises because the mass of the compound system is the sum of the energies of the constituents. This factor cancels 
the factor $1/E$ in the volume element, just as in the calculations of Stoner and Chandrasekhar. }
$$
\hat\rho =  q^4 ={1\over2\pi^2}\int E \,{d^4q\over E},~~ \hat p =\int q^4d\psi =
{1\over 2\pi^2}\int q \,{ d^4q\over E}.\eqno(5.2)
$$
 The equation of state   that  has been developed here is consistent with the observed velocity  distribution, even for the innermost  orbiters, but  this does not give enough information to develop a microscopic model of dark matter.

It is essentially a   hydrodynamical system. At very low densities the star is an $ n = 4$
polytrope, 
$$
 \hat p = {1\over5}\, \hat \rho^{5/4}, ~~ \hat \rho = \sinh^4\psi.
 $$
 With 
 $$
 \rho =\alpha \hat\rho, ~~\alpha = {16 k\over b^2 w} = 6.1 \times 10^{-24},
 $$
  and 
 $$
 p =\alpha\beta \hat p,~~ \beta =  c^2 k  = 9.4 \times 10^{14},
 $$
 it works out to
 $$
 p = A \rho^{5/4}, ~~ A = {1\over 5}\alpha^{-1/4}\beta = 6.0 \times 10^{20}.
 $$
 At very high densities the adiabatic index is effectively   infinite, $\hat p = \hat\rho/4$ and
 $$
  p = {\beta\over 4} \rho = 2.35\times 10^{14}\rho.
$$ 
 This relation is confirmed at  the point where $\psi = 10.48$,    
 $\rho = .00122 g/cm $, where we are in the regime of high densities.

\b\b

\no{\bf VI. Theory}

The reader will have noticed that the equations that were used are classical,
but a closer look reveals some novelties.
\b
1. The integrated hydrostatic condition is classical; a short calculation shows that
taking the  gradients of both sides leads to the usual hydrostatic condition,
$$
\rho\,{\rm grad}\, \phi = -{\rm grad \,p}.
$$
The difference is that the integrated form incorporates the boundary condition that
fixes the speed of light at infinity.     

2. It is of interest to ask why this boundary condition has not been applied previously.
One part of the answer is that  the stars of Eddington and Chandrasekhar  all have abrupt boundaries at a point where the temperature and the density vanish, so that these fields are not continuous.
Another part of the reason is that the equations employed by Eddington (1926),  
Chandrasekhar (1935) and many others
differ in one particular from ours: The factor $\e^{-\nu}$ that multiples the density
in Eq.(3.8) is absent; consequently, the function $\nu$  is represented only by its derivative,  so that
fixing its boundary value has no meaning. 

In view of this difference we need to justify our approach. All equations used are variational equations based on the following action,
$$
A = {1\over 8\pi G}\int d^4x \sqrt{-g} R +
 \int d^4x \sqrt{-g}\Big(\rho(g^{\mu\nu}\Psi_{,\mu}\Psi_{,\nu} - c^2) - f(\rho)\Big).
 $$
The non relativistic approximation of the matter lagrangian is one that was used
by Fetter and Walecka (1980) to obtain a variational formulation of hydrodynamics: the equation of continuity and the Bernoulli equation (in integrated form).\footnote{$^5$}{The non relativistic approximation is taken by setting $\Psi = c^2 t + \Phi$, neglecting $1/c^2$
terms and interpreting $\Phi$ as the velocity potential.}  
The term $f(\rho)$ is the  free energy, as is seen from the structure of the equations of motion
(Fr\o nsdal 2007, 2008). The variational approach to hydrodynamics is an application of the Gibbs variational principle to the case that the temperature is frozen, so that neither temperature nor entropy plays any role, as is appropriate for a treatment of dark matter,
effectively a hydrodynamical system. The density $\rho$ is denoted $\rho + p$
by Eddington; this is a matter of notation, and irrelevant in the 
immediate context, since $p/c^2 \rho$  never exceeds $10^{-4}$. The 
gradient of the field $\Psi$ corresponds to Tolman's vector field $U$; for a stationary solution $\Psi_{,0}$ is a constant, while Tolman's normalization condition leads to
$U_0 = \sqrt{g_{00}}$. This is what gives rise to the cancellation of this metric function
in Tolman's equations of motion, and it constitutes an important difference in principle between our approach and that of Tolman.\footnote{$^6$}{ In a variational approach
it is important to specify the independent variables; any constraint  is the source of great complications.}

3. Another consequence of action principle dynamics is that the current is conserved,
$$
\p_\mu J^\mu = 0,~~ J^\mu =  g^{\mu\nu}\Psi_{,\nu} \rho.
$$
The usual approach   does not admit a conserved current
and breaks with non relativistic theory in this respect.

Although the current is conserved, it is not directly related to the ``mass" as we would define it. Our approach  to stellar structure is to start the analysis from the outside,
using observational data. In the models considered here the metric has the asymptotic form
$$
c^{-2}g^{00} = 1 -+{2MG\over r},
$$
with $M$ constant. This is an obervational datum, measured by observing the motion of test bodies. In this paper, that is what we call the  mass of the star.  Now it is always pointed out that  the function $m$ that appears in (3.4) tends to $M$ at infinity.
 In the traditional approach  the equation is just $dm/dr = wr^2
 \rho$ and 
 the mass can be expressed as an integral over the density, provided that $m$ vanishes at the origin.    \footnote{$^7$}{It has   
been pointed out that the integral $\int\rho d^3 r$ does not have the correct measure. Kippenhahn and Weigert (1990) calls this situation hazardous, but no one seems to have taken the warning seriously.}  
Our theory preserves the continuity equation of classical hydrodynamics, but there is no direct connection between mass and the conserved quantity.  \footnote{$^8$}{The non relativistic gravitational potential
 arises entirely from the time component of the metric, while the function $m$ is in the space component. The unfortunate association of this function with the mass is due to Tolman's normalization condition.}  Consequently, there is no imperative to postulate that the function $m$ vanishes at the origin.

\b
 An equation of state in hydrodynamics  is a relation between density and pressure.
The pressure term in Einstein's equation is not as important as the role that is played by the pressure in the hydrostatic equation.  In our approach  the equation of state follows from the expression that is chosen  for the free energy density. Our approach preserves all the structure of hydrodynamics, including   the equation of continuity.
\b
Ultimately, the model of our Galaxy must be improved by including the 
contribution of visible matter;  the overall structure can be studied 
in terms of an idealized, continuous, spherically symmetric distribution
that makes an important, additional  contribution to the action.
The absence of any interaction between dark and visible matter makes this
very straightforward; in the absence of any interaction between the two kinds of matter the free energy density is additive. Note that it is
essential to recognize the roles of two quite different  density fields.

\b\b
\no{\bf VII. Speculating about the center}

In this paper the study of our galaxy has been approached from the outer regions,
because that is where observations have been made up to this time. The analysis is
especially interesting because nothing is known about the nature of dark matter.
Consequently, there is no need to ask what pressures and densities can be
allowed; there is no way that we can answer  questions of this kind.

Historically, a number of statements have been made that would place limits on the mass of certain types of stars. These statements all rely on two assumptions:\break 
(a) that the nature of the matter within the star is within the limits of our knowledge
and (b)  that the total mass is  the integral of the function $m$, from the center outwards.
If these premises may be said to be  reasonable as  far as stars made up of ordinary matter is concerned, they cannot be applied with any degree of confidence to the
new situation that is faced in connection with dark matter. The bump in the density
is a common feature of relativistic models; here it is predicted to occur at a 
very small distance  from the center. It is conceivable that future observation
may validate this prediction, but what happens inside is beyond our reach.

 \b\b
\no{\bf Acknowledgements}

We are grateful to  Robert Delbourgo, Paul Frampton, Philip Mannheim, John Moffat and  Abhishek Pathak for stimulating conversations.
 
\b
\no{\bf References}

\no Chandrasekhar, S.,   Month.Not. R.A.S. {\bf 95} 222 (1935).  
 
\no Chandrasekhar, S., {\it An introduction to stellar structure}, U. Chicago press 1938.
 
\no  Delbourgo, R. and Lashmar, D.   ``Born Reciprocity and the  $1/r$  potential", 

Found. of Phys. 38, 995-1010, 2008.

\no Emden,  J. R., {\it Gaskugeln}, Teubner 1907.

\no Eddington, A.S.,    {\it The internal constitution of stars}, Dover. N.Y., 1959.

\no Fetter, A.L. and  and Walecka, J.D., {\it Theoretical mechanics of particles and continua},  

McGraw-Hill, N.Y. 1980.

 \no Fr\o nsdal,    C. ``Heat and Gravity. I. The action principle",   arXiv:0812.4990. 

 \no Fr\o nsdal,  C.  ``Ideal stars and General Relativity", Gen.Rel.Grav. {\bf 39} 1971-2000 (2007).
 
\no Ghez, A. M., Salim, S., Weinberg, N. N., Lu, J. R., Do, T., Dunn, J. K., Matthews, K., 

Morris,
M., Yelda, S., Becklin, E. E., Kremenek, T., Milosavljevic, M., Naiman, J., 

 ÒMeasuring
Distance and Properties of the Milky WayÕs Central Supermassive Black 
 
 Hole with Stellar
Orbits,Ó ApJ, 689, 1044 (2008).
 
\no Gibbs, J.W., ``On the equilibrium of heterogeneous substances",

Trans.Conn.Acad.  108-248 (1878).

\no Hartle, J.B. ``Bounds on the mass and moment of inertia of non-rotating neutron stars", 

 Physics Reports {\bf 46} 201-247 (1978).

\no Kippenhahn,  R. and Weigert, A., {\it  Stellar structure and evolution}, Springer-Verlag  1990.

\no Landau, L. D. and Lifshitz, E. M., {\it Statistical physics}, Pergamon Press 1958.	 

\no Mannheim, P., ``Making the case for conformal gravity", arXiv 1101.2186.

\no Oppenheimer, J.R. and Volkov, G.M., ``On massive neutron stars", 

Physical Review {\bf 55} 374-381 1939).

\no Stoner, E.C.,  ``The minimum pressure of a degenerate electron gas",  

Monthly Notices R.A.S., {\bf 92} 651-661 (1932).

\no Tolman,  R.C., {\it Thermodynamics and Cosmology}, Clarendon, Oxford 1934.  

  \end
 
  The equations become more complicated but they can still be solved analytically. But this cannot be supported by observations. 
At this time there is no experimental information about the center of the Great Attractor; 
consequently it is not important to strain the model by extending the integration
much further in that direction. In particular, the integral of the function $m$, 
from the center outwards, contains no  information that is useful at this time.

\b

Other numerical results are as follows. The density is positive; there is a characteristic bump in the density 
profile  - see Fig.3b - where the density reaches the highest value, $ \rho =  100
g/cm^3$ at $x = 21.6$,
$$
\rho_{\rm max} = 100g/cm^3 ~~{\rm at}~~ r =   2.4\times 10^{9}.
$$
This ``object" is comparable to our Sun, in size, mass and gravitational field strength.
  At the shortest distance observed for an orbiting satellite, 
$r = 2\times10^{15} ~ (x = 35.23)$, the density is about $ 1.3\times10^{-10}$. The
pressure has a similar profile - Fig.4, with a peak value of  $2.5\times 10^{16}$.  The  gravitational field $-\phi(x) = c^{-2}g_{00}-1$ also has a maximum  -see Fig.5, reaching a maximum value of $\phi = .00003 $ at the same point. This is about  ten times stronger than the gravitational potential at the surface of the Sun. 
 The appearance of such  shapes is very common when 
polytropic equations of state are used; they are relativistic features not seen in the weak field approximation. For some stars the maximum value
of $\phi$ can rise to get very close to the limiting value of unity, at which point a horizon would appear.

\vskip3.5cm

\parindent=1pc

\vskip-3cm
 
\epsfxsize.6\hsize
\centerline{\epsfbox{Fig.3.eps}}
\vskip-.1cm

Fig.3. A plot of $\ln\rho/ \ln  r$ against $ x = \ln r$.
 \vskip2cm
\vskip2cm

\epsfxsize.6\hsize
\centerline{\epsfbox{Fig.4a.eps}}
\vskip0cm

Fig.4a. The characteristic, inner  density profile plotted against $x = \ln r$, with the peak at $r = 9.67 \times 10^{12} cm$.
\ve
\vskip3.5cm

 \vskip-3.2cm
 
\epsfxsize.6\hsize
\centerline{\epsfbox{Fig.1b.eps}}
\vskip0cm

Fig.5a. The pressure profile in the inner region.

\vskip3.5cm

\vskip-3cm
 
\epsfxsize.6\hsize
\centerline{\epsfbox{Fig.1b.eps}}
\vskip0cm

Fig.6. The potential has a maximum at the same point. The density and pressure peaks are narrower because of the high value of the ``polytropic index".

\b
\b

\parindent=1pc
\vskip3cm
 
\vskip-3.cm
 
\epsfxsize.6\hsize
\centerline{\epsfbox{Fig.2b.eps}}

Fig. 7. The velocity distribution predicted by the relativistic equations of motion, in km/sec.   Compare Fig.1. 
 \end